\documentclass[11pt]{article}
\begin{document}

\begin{center}
\large \textbf{Overcharging problem and thermodynamics in five-dimensional black holes}
\end{center}

\begin{center}
Deyou Chen and Shan Zeng
\end{center}

\begin{center}
School of Science, Xihua University, Chengdu 610039, China

E-mails: deyouchen@hotmail.com, shanzengxhu@hotmail.com
\end{center}

{\bf Abstract:} In this paper, the overcharging problem and thermodynamics in the extended phase spaces of the five-dimensional spherically symmetric topological black holes are investigated by absorptions of scalar particles and fermions. The cosmological constant is regarded as a variable related to pressure and its conjugate quantity is a thermodynamic volume. The first law of thermodynamics is recovered. The second law is violated in the extended phase space of the extremal and near-extremal black holes. The overcharging problem is tested by the existence of the event horizons. The event horizon is determined by the metric component $f(r)$. The minimal values of the metric component at the final stage show that the extremal and near-extremal black holes can not be overcharged.

\section{Introduction}

It is well known that a spacetime singularity can be formed at the end of gravitational collapse. Near the singularity, physical laws fail. To avoid the physical damage caused by the singularity, Penrose pointed out that under normal material properties and initial conditions, naked singularities cannot be formed in the real physical process \cite{RP}. This is the weak form of the cosmic censorship conjecture. It shows that singularities are not naked and hide behind event horizons. This cosmic censorship conjecture plays an important role in black hole physics, and its validity is a necessary condition to ensure the predictability of physical laws. Although its correctness is widely accepted, there is no complete evidence to prove it.

An effective method to test the weak cosmic censorship conjecture (WCCC) is the Gedanken experiment designed by Wald \cite{RMW}. In this experiment, a particle with energy, large enough angular momentum and charge is thrown into a charged rotating black hole to test the evolution of the black hole. The evolution result is described by the new spacetime metric. If the final state of the black hole evolution is no longer a black hole, there is no event horizon and the singularity is naked. The solution of the event horizon is determined by the metric components. Therefore, one only needs to check the existence of the horizon to judge whether the singularity is naked. The energy, charge and angular momentum of the black hole can also be gotten by the interaction between the black hole and a field. The field has a finite energy (i.e. it contains a wave packet). At first, the field does not exist, there is only the black hole. Then, the field comes from infinity and interacts with the black hole. Due to the interaction, energy, charge and angular momentum are transferred between the black hole and field. Part of the field is absorbed by the black hole and the other part is reflected back to infinity. Finally, the field decays, and the energy, charge and angular momentum of the black hole change. Using this field view, Semiz first adopted a complex scalar field to test the WCCC and found that the dyonic Kerr-Newman black hole could not be overcharged \cite{IS}. This result was supported by the work of Toth \cite{GZT}. Based on the seed work of Wald, the validity of the WCCC in the various spacetimes has been tested \cite{CO,MS,SH,RS1,JS,SS,IST,SH2,GZ,DS1,HMS,KD1,KD2,HS,RV,FGS,BG1,GG,BG6,CHS,YW,LWL,LWL1,LWL2}. In the relevant discussions, the self-force effect, tunnelling effect, radiation reaction were taken into account \cite{CB,CBSM,BCK1,ZVPH}. However, there is no consensus. In these researches, the black holes are assumed to exist in advance. The formation of the naked singularity was also discussed by the numerical method \cite{MM}. In the discussion, it was assumed that a celestial body before particles being absorbed is not a black hole, but a general dense star.

The cosmological constant used to be a fixed constant. Recently, it was seen as a variable in the work of Kastor and others \cite{KRT1,KRT2,LPVP,KM1}. One of the reasons for this consideration is that the cosmological constant, as a variable, can reconcile the inconsistency between the first law of thermodynamics of black holes and the Smarr relation derived from the scaling method. The second reason is that if people consider "more fundamental" theories, physical constants, such as the gauge coupling constant, Yukawa coupling constant, Newtonian constant or cosmological constant arise as vacuum expectation values are not fixed in advance and can vary \cite{CM}. The cosmological constant was seen as a pressure, and its conjugate quantity is a thermodynamic volume. Using the standard thermodynamics, Kubiznak and Mann studied the critical behavior of the charged anti-de Sitter black hole. They found that its critical index was consistent with that of van der Waals system \cite{KM1}. The original first law of black hole thermodynamics was modified to

\begin{eqnarray}
dM=TdS+\phi dQ+VdP+\Omega dJ.
\label{eq1.1}
\end{eqnarray}

\noindent The corresponding Smarr relation is

\begin{eqnarray}
(n-3)M=(n-2)TS+ (n-2)\Omega J+ (n-3)\phi Q -2VP.
\label{eq1.2}
\end{eqnarray}

\noindent where the mass $M$ is expressed as the enthalpy and $n$ is the dimension of the spacetime. Introducing the thermodynamic pressure and volume into the black hole system, one can use the standard thermodynamic formulae to calculate various thermodynamic quantities and discuss various thermodynamic phenomena. In the extended phase space, many interesting thermodynamic phenomena have been found, including the reentant phase transition, repulsive interaction in the high temperature black holes, multiphase transition in the quasi topological gravity and Lovelock gravity, and other thermodynamic phenomena \cite{FKMS,HBM,CCLY,WLM,WL,ZZW1,ZZW2}.

In the recent work, Gwak studied the WCCC in the extended phase space of the rotating black hole, where the pressure and volume were introduced \cite{BG}. He found that the black hole could not be overspun by throwing a particle with energy and large enough angular momentum into the hole. This result is in consistence with that gotten in the new version Gedanken experiment \cite{SW1,RMW2,ASZZ1,CLN,MS2,HJ}. In the extremal and near-extremal cases, the entropies decreased. This means the violation of the second law of thermodynamics in the extended phase space. The same result was gotten in the subsequent work \cite{BG5,ZZ1,HMT,WWY1,WWY2,WWY3,HZH,ZZ2}. In these work, an important relation between the momentum, energy and angular momentum or charge of the particle was gotten by the Hamilton-Jacobi equation.

In this paper, we investigate the WCCC and thermodynamics in extended phase spaces of five-dimensional charged topological black holes by absorptions of scalar particles and fermions. These black holes are a spacial case of a $(n + 1)$-dimensional Einstein-Maxwell-dilaton gravity solution \cite{AS}, where the dilaton $\alpha = 0$. The test particles in this paper are scalar particles and fermions, which satisfy the Klein-Gordon equation and Dirac equation in curved spacetime, respectively. The first law of thermodynamics is recovered by the absorptions of the particles. In the extremal and near-extremal cases, the second law is violated in the extended phase spaces. The overcharging problem is tested by the existence of the event horizon. The minimal values of the metric component $f(r)$ at the final stages show that the extremal and near-extremal black holes can not be overcharged.

The rest is organized as follows. The five-dimensional topological black hole is given and the thermodynamics are discussed in the next section. In section 3, we investigate the absorptions of the scalar particle and fermion by the black hole. The relation between the momentum, energy and charge of the particle is gotten. In section 4, the thermodynamics in the extended phase space are investigated by the absorptions of the particles. In section 5, the overcharging problem is tested by throwing a particle in the near-extremal and extremal black holes. The last section is devoted to our discussion and conclusion.

\section{The black hole solution}

The action described the $(n + 1)$-dimensional Einstein-Maxwell-dilaton gravity is \cite{CHM}

\begin{eqnarray}
\mathcal{S} = \frac{1}{16\pi}\int{dx^{n+1}\sqrt{-g}\left[\mathcal{R} - \frac{4}{n-1}\left(\nabla \Phi\right)^2-V(\Phi) - e^{-\frac{4\alpha \Phi}{n-1}}F_{\mu\nu}F^{\mu\nu}\right]},
\label{eq2.1}
\end{eqnarray}

\noindent where $n\ge 3$, $\mathcal{R}$ is the Ricci scalar curvature, $\Phi$ is a dilaton field, and $V(\Phi)$ is a potential for $\Phi$,

\begin{eqnarray}
V(\Phi) = 2 \Lambda e^{\frac{4\alpha \Phi}{n-1}} + \frac{k(n-1)(n-2)\alpha^2}{b^2(\alpha^2-1)}e^{\frac{4 \Phi}{(n-1)\alpha}}.
\end{eqnarray}

\noindent $\Lambda$ is a negative cosmological constant. $\alpha$ is a constant determining the coupling strength of the scalar and electromagnetic fields. $F_{\mu\nu}=\partial_{\mu}A_{\nu}-\partial_{\nu}A_{\mu}$ is the electromagnetic field tensor and $A_{\mu}$ denotes the electromagnetic potential. The black hole solutions in the Einstein-Maxwell-dilaton gravity with the dilaton coupling parameter $\alpha$ are given by \cite{AS}

\begin{eqnarray}
ds^{2}=-f(r)dt^{2}+\frac{dr^{2}}{{f(r)}}+r^{2}R^{2}(r)d\Omega^{2}_{k,n-1},
\label{eq2.2}
\end{eqnarray}

\noindent where

\begin{eqnarray}
f(r) &=& -\frac{k(n-2)(\alpha^2 +1)^2 b^{-2\gamma} r^{2\gamma} }{(\alpha^2-1)(\alpha^2+n-2) }-\frac{m}{r^{(n-1)(1-\gamma)-1}} \nonumber\\
&&+\frac{2q^2(\alpha^2 +1)^2 b^{-2(n-2)\gamma} r^{2(n-2)(\gamma-1)} }{(\alpha^2-1)(\alpha^2+n-2) } +\frac{2\Lambda (\alpha^2 +1)^2 b^{2\gamma} r^{2(1-\gamma)} }{(n-1)(\alpha^2-n) },
\label{eq2.3}
\end{eqnarray}

\noindent $R(r) = e^{\frac{2\alpha \Phi}{n-1}}$, $\Phi (r)=\frac{(n-1)\alpha}{2(1+\alpha^2)}ln\left(\frac{b}{r}\right)$, $\gamma = \frac{\alpha^2}{\alpha^2+1}$ and $b$ is an arbitrary constant. $d\Omega^{2}_{k,n-1}$ describes a $(n-1)$-dimensional subspace with $n(n-1)k$ constant curvature and volume  $\omega_{n-1}$. $k$ is a constant and characterizes the hypersurface. Without loss of generality, the value of $k$ is $-1, 0, 1$, which correspond to a negative (hyperbolic), zero (flat), positive (elliptic) constant curvature hypersurface in the metric (\ref{eq2.2}), respectively. $m$ and $q$ are integration constants. According to the definition of mass due to Abbott and Deser \cite{AD1,AD2}, the mass takes the form

\begin{eqnarray}
M=\frac{(n-1)b^{(n-1)\gamma} m\omega_{n-1}}{16\pi (\alpha^2 +1) }.
\label{eq2.4}
\end{eqnarray}

\noindent The charge is

\begin{eqnarray}
Q = \frac{q\omega_{n-1}}{4\pi}.
\label{eq2.5}
\end{eqnarray}

\noindent The electric potential is $A_t = -\frac{qb^{(3-n)\gamma}}{\lambda r^{\lambda}}$, where $\lambda = (n-3)(1-\gamma)+1$. The thermodynamic properties in the extended phase space of the $(n + 1)$-dimensional dilaton black holes were deeply discussed in \cite{ZZMZ,DKS,HSPP,DSM,HSPP1,HSPP2}.

In this paper, our interest is focused on the overcharging problem and thermodynamics in the extended phase spaces of the five-dimensional black holes. Therefore, we let $n=4$ in the metric (\ref{eq2.2}). Considering that the cosmological constant is a variable related to pressure, $P= -\frac{\Lambda}{8\pi}=\frac{n(n-1)}{16\pi l^2}$, we replace the cosmological constant with the pressure in the following discussions. In the absence of the non-trivial dilaton ($\alpha =0$), the solution (\ref{eq2.3}) is reduced to

\begin{eqnarray}
f(r) = k-\frac{M}{r^2} \frac{16\pi}{3\omega_3}+\frac{Q^2}{r^4} \frac{16\pi^2}{3\omega_3^2}+\frac{4}{3}\pi r^2P,
\label{eq2.6}
\end{eqnarray}

\noindent Thus the metric (\ref{eq2.2}) describes a five-dimensional anti-de Sitter topological spacetime where $R(r) = 0$. Now the electric potential is given by

\begin{eqnarray}
A_{t} = -\frac{2\pi Q}{\omega_3 r^2}.
\label{eq2.7}
\end{eqnarray}

\noindent The Hawking temperature and entropy are

\begin{eqnarray}
T = \frac{f^{\prime} (r_+)}{4\pi}, \quad \quad S=\frac{\omega_3 r_+^3}{4},
\label{eq2.7}
\end{eqnarray}

\noindent respectively. $r_+$ is the event horizon radius derived by $f(r)=0$. The mass is expressed by the entropy, charge and pressure

\begin{eqnarray}
M = \frac{3k \omega_{3}}{16\pi}\left(\frac{4S}{\omega_3}\right)^{\frac{2}{3}} +\frac{\pi Q^2}{\omega_{3}} \left(\frac{\omega_3}{4S}\right)^{\frac{2}{3}} +\frac{\omega_{3} P}{4} \left(\frac{4S}{\omega_3}\right)^{\frac{4}{3}}.
\label{eq2.8}
\end{eqnarray}

\noindent Here, $S$, $Q$ and $P$ are regarded as the extensive parameters for the mass $M$. Their conjugate quantities are the temperature, electric potential and volume, respectively.

\begin{eqnarray}
T &=& \left(\frac{\partial M}{\partial S}\right)_{Q,P}=\frac{k\omega_3}{8\pi}\left(\frac{4}{\omega_3}\right)^{\frac{2}{3}}S^{-\frac{1}{3}}-\frac{2\pi Q^2}{3\omega_3}\left(\frac{\omega_3}{4}\right)S^{-\frac{5}{3}}+\frac{\omega_3 P}{3}\left(\frac{4}{\omega_3}\right)^{\frac{4}{3}}S^{\frac{1}{3}}, \nonumber\\
\phi &=& \left(\frac{\partial M}{\partial Q}\right)_{P,S}=\frac{2\pi Q}{\omega_3 r_+^2}, \quad\quad V = \left(\frac{\partial M}{\partial P}\right)_{S,Q}=\frac{\omega_3 r_+^4}{4}.
\label{eq2.9}
\end{eqnarray}

\noindent After a straightforward calculation, it is easily found that these thermodynamic quantities obey the first law of thermodynamics

\begin{eqnarray}
dM=TdS+\phi dQ+VdP,
\label{eq2.10}
\end{eqnarray}

\noindent and the Smarr relation

\begin{eqnarray}
2M =3TS+2\phi Q -2VP.
\label{eq2.11}
\end{eqnarray}

\noindent In what follows the thermodynamics are investigated by the absorptions of the scalar particle and fermion and the first law of thermodynamics is recovered.

\section{Particles' absorption}

\subsection{Scalar particle's absorption}

In this subsection, we discuss the absorption of the scalar particle in the five-dimensional spacetime. In the curved spacetime, the motion of the charged scalar particle obeys the Klein-Gordon equation

\begin{eqnarray}
\frac{1}{\sqrt{-g}}\left(\frac{\partial}{\partial  x^{\mu}}-\frac{iq_e}{\hbar}A_{\mu}\right)\left[\sqrt{-g}g^{\mu\nu}\left(\frac{\partial}{\partial x^{\nu}}-\frac{iq_e}{\hbar}A_{\nu}\right)\right]\Psi_{S}-\frac{m_e^{2}}{\hbar^{2}}\Psi_{S}=0,
\label{eq3.1}
\end{eqnarray}

\noindent where $m_e$ and $q_e$ are the mass and charge of the particle, respectively. $\Psi_{S}$ is the scalar field. Following the ansatz \cite{GRS,EGLSY,SG}, we use the $WKB$ approximation and adopt the following expression of the wave function

\begin{eqnarray}
\Psi_{S}=exp(\frac{i}{\hbar}I_s+I_{1}+\mathcal{O}(\hbar)).
\label{eq3.2}
\end{eqnarray}

\noindent Inserting the above ansatz and the contravariant metric components of the five-dimensional black hole into the Klein-Gordon equation yields an equation about the action $I_s$. Dividing by the exponential term and taking the leading contribution of $\hbar$, we get

\begin{eqnarray}
f^{-1}(\partial_{t}I_s-q_e A_t)^2-f(\partial_{r}I_s)^2-g^{22}(\partial_{\theta_2}I_s)^2-g^{33}(\partial_{\theta_3}I_s)^2-g^{44}(\partial_{\theta_4}I_s)^2+m_e^2=0.
\label{eq3.3}
\end{eqnarray}

\noindent Here we use $\theta_2$, $\theta_3$, $\theta_4$ to express the coordinates $x^2$, $x^3$, $x^4$, respectively. And $g^{22}$, $g^{33}$ and $g^{44}$ are the corresponding contravariant metric components in the five-dimensional metric. Their expressions are not specified here, since they do not affect the later calculation result. This is different from the previous discussion of tunneling radiation with consideration of quantum gravity effects \cite{CJWY}. Taking into account the symmetries of the spacetime, we carry out the separation of variables to the action,

\begin{eqnarray}
I_s=-\omega t+W(r)+\Theta(\theta_2,\theta_3,\theta_4).
\label{eq3.4}
\end{eqnarray}

\noindent In the above equation, $\omega$ is the energy of the absorbed scalar particle. Substituting the separated action into Eq. (\ref{eq3.3}) and solving it yield

\begin{eqnarray}
\partial_{r}W=\pm\frac{\sqrt{\left(\omega - q_e \frac{2\pi Q}{\omega_3 r^2}\right)^{2}+\left[m_e^{2}- g^{22}(\partial_{\theta_2}\Theta)^{2}-g^{33}(\partial_{\theta_3}\Theta)^{2}
-g^{44}(\partial_{\theta_4}\Theta)^{2}\right]f}}{f},
\label{eq3.5}
\end{eqnarray}

\noindent where $+ (-)$ denote the cases of the ingoing (outgoing) particles. We assume that the particle is completely absorbed by the black hole. Then the negative sign in the above equation is neglected. Defining $p_r=\partial_r I_s$, we get

\begin{eqnarray}
p^{r} &=& g^{rr}p_{r} \nonumber\\
&=& \sqrt{\left(\omega - q_e \frac{2\pi Q}{\omega_3 r^2}\right)^{2}+\left[m_e^{2}- {g^{22}}(\partial_{\theta_2}\Theta)^{2}-{g^{33}}(\partial_{\theta_3}\Theta)^{2}-{g^{44}}(\partial_{\theta_4}\Theta)^{2}\right]f},
\label{eq3.6}
\end{eqnarray}

\noindent Near the event horizon, we have $f (r)\to 0$ and the above equation is simplified to

\begin{eqnarray}
p^{r}=\omega-q_e\phi,
\label{eq3.7}
\end{eqnarray}

\noindent where $\phi=\frac{2\pi Q}{\omega_3 r_+^2}$ represents the electric potential at the event horizon. The condition of the superradiation is that the boundary condition of the scalar field should be in the asymptotic region and $\omega<q_e\phi$. In this paper, the effects of superradiation are not taken into account, which leads to $\omega>q_e\phi$ and $p^r>0$. The relation (\ref{eq3.7}) plays an important role in the derivation of thermodynamics and is recovered by the absorption of the fermion in the next subsection.

\subsection{Fermion's absorption}

In this subsection, we use a charged fermion with spin-1/2 to discuss its absorption at the event horizon. The motion of the fermion obeys the Dirac equation in curved spacetime

\begin{equation}
i\gamma^{\mu}\left(\partial_{\mu}+\Omega_{\mu}+\frac{i}{\hbar}q_{e}A_{\mu}\right)\Psi_{F}+\frac{m_{e}}{\hbar}\Psi_{F}=0,
\label{eq4.1}
\end{equation}

\noindent where $m_{e}$ and $q_e$ denote the mass and charge of the fermion, respectively. $\Psi_{F}$ is the wave function. $\Omega _\mu \equiv\frac{i}{2}\omega _\mu\, ^{a b} \Sigma_{ab}$, $\omega _\mu\, ^{ab}$ is the spin connection defined by the ordinary connection and the tetrad $e^\lambda\,_b$, namely, $\omega_\mu\,^a\,_b=e_\nu\,^a e^\lambda\,_b \Gamma^\nu_{\mu\lambda}-e^\lambda\,_b\partial_\mu e_\lambda\,^a$. The Greek indices are raised and lowered by the curved metric $g_{\mu\nu}$. The Latin indices are governed by the flat metric $\eta_{ab}$. The construction of the tetrad needs to use the following definitions

\begin{equation}
g_{\mu\nu}= e_\mu\,^a e_\nu\,^b \eta_{ab},\hspace{5mm} \eta_{ab}=
g_{\mu\nu} e^\mu\,_a e^\nu\,_b, \hspace{5mm} e^\mu\,_a e_\nu\,^a=
\delta^\mu_\nu, \hspace{5mm} e^\mu\,_a e_\mu\,^b = \delta_a^b.
\label{eq4.3}
\end{equation}

\noindent The Lorentz spinor generators are defined by $\Sigma_{ab}= \frac{i}{4}\left[ {\gamma ^a ,\gamma^b} \right]$ and $\gamma^a$ satisfy $\{\gamma ^a ,\gamma^b\}= 2\eta^{ab}$. We can readily construct the $\gamma^\mu$'s in curved spacetime as

\begin{eqnarray}
\gamma^\mu = e^\mu\,_a \gamma^a, \hspace{7mm} \left\{ {\gamma ^\mu,\gamma ^\nu } \right\} = 2g^{\mu \nu }. \label{eq4.5}
\end{eqnarray}

For a spin-$1/2$ fermion, there are two states corresponding to spin up and spin down. First of all, we discuss the spin up state. The wave function takes the form

\begin{eqnarray}
\Psi_{F\uparrow}=\left(\begin{array}{c}
A\\
0\\
B\\
0
\end{array}\right)\exp\left(\frac{i}{\hbar}I_{\uparrow}\right),
\label{eq4.6}
\end{eqnarray}

\noindent where $A$, $B$ and $I_{\uparrow}$ are the functions of $(t,r,\theta_2,\theta_3,\theta_4)$. To solve the wave function, we should construct the $\gamma^\mu$ matrices. There are many choices to construct them. For the five-dimensional metric, we first choose the tetrad

\begin{eqnarray}
e_\mu\,^a = \rm{diag}\left(\sqrt f, 1/\sqrt f, \sqrt{g^{22}},\sqrt{g^{33}},\sqrt{g^{44}}\right).
\end{eqnarray}

\noindent Then the $\gamma^\mu$ matrices are

\begin{eqnarray}
\gamma^{t}=\frac{1}{\sqrt{f\left(r\right)}}\left(\begin{array}{cc}
0 & I\\
-I & 0
\end{array}\right), &  & \gamma^{\theta_2}=\sqrt{g^{22}}\left(\begin{array}{cc}
0 & \sigma^{1}\\
\sigma^{1} & 0
\end{array}\right),\nonumber \\
\gamma^{r}=\sqrt{f\left(r\right)}\left(\begin{array}{cc}
0 & \sigma^{3}\\
\sigma^{3} & 0
\end{array}\right), &  & \gamma^{\theta_3}=\sqrt{g^{33}}\left(\begin{array}{cc}
0 & \sigma^{2}\\
\sigma^{2} & 0
\end{array}\right), \nonumber \\
\gamma^{\theta_4}=\sqrt{g^{44}}\left(\begin{array}{cc}
-I & 0\\
0 & I
\end{array}\right).
\label{eq4.7}
\end{eqnarray}

\noindent $\sigma^{1}$, $\sigma^{2}$ and $\sigma^{3}$ are the Pauli matrices and $I$ is a identity matrix. Inserting the spin connection, wave function and gamma matrices into the Dirac equation, dividing the exponential term and multiplying by $\hbar$ yield four equations to the leading order in $\hbar$. They are

\begin{eqnarray}
-B\frac{1}{\sqrt{f}}\left({\partial_{t}I_{\uparrow}-q_eA_{t}}\right)-B\sqrt{f}\partial_{r}I_{\uparrow}+ A\sqrt{g^{44}}\partial_{\theta_4}I_{\uparrow}+Am_{e}=0,
\label{eq4.8}
\end{eqnarray}

\begin{eqnarray}
A\frac{1}{\sqrt{f}}\left({\partial_{t}I_{\uparrow}-q_eA_{t}}\right)-A\sqrt{f}\partial_{r}I_{\uparrow}- B\sqrt{g^{44}}\partial_{\theta_4}I_{\uparrow}+Bm_{e}=0,
\label{eq4.8}
\end{eqnarray}

\begin{eqnarray}
-B\left[{\sqrt{g^{22}}\partial_{\theta_2}I_{\uparrow}+i\sqrt{g^{33}}\partial_{\theta_3}I_{\uparrow}}\right]=0,
\label{eq4.10}
\end{eqnarray}

\begin{eqnarray}
-A\left[{\sqrt{g^{22}}\partial_{\theta_2}I_{\uparrow}+i\sqrt{g^{33}}\partial_{\theta_3}I_{\uparrow}}\right]=0.
\label{eq4.11}
\end{eqnarray}

\noindent The last two equations are simplified to an equation and yield $g^{22}(\partial_{\theta_2}I_{\uparrow})^{2}+g^{33}(\partial_{\theta_3}I_{\uparrow})^{2}=0$. Previous discussions have shown that the contribution of the angular part does not affect the results of the tunneling radiation when quantum gravity effects are not taken into account \cite{CJWY}. Therefore, we focus on the first two equations. Due to the same symmetry in the black hole spacetime, we still adopt the separation of variables in the above subsection

\begin{eqnarray}
I_{\uparrow}=-\omega t+W(r)+\Theta(\theta_2,\theta_3,\theta_4).
\label{eq4.12}
\end{eqnarray}

\noindent Now $\omega$ denotes the energy of the absorbed fermion. Combining these two equations and eliminating $A$ and $B$, we get an equation about the action $I_{\uparrow}$. Substituting the separated variables into this equation and solving it, we get

\begin{eqnarray}
\partial_{r}W=\pm\frac{\sqrt{\left(\omega - q_e \frac{2\pi Q}{\omega_3 r^2}\right)^{2}+\left[m_e^2-g^{44}\partial_{\theta_4}I_{\uparrow}\right]f}}{f},
\label{eq4.13}
\end{eqnarray}

\noindent where $+ (-)$ denote the cases of the ingoing (outgoing) fermion. As addressed in the above subsection, the particle is assumed to be completely absorbed. Therefore, there is no outgoing fermion. Using the definition $p_{r}=\partial_{r}I_{\uparrow}$, we get $p^{r}=g^{rr}\partial_{r}I_{\uparrow}$. Near the event horizon,

\begin{eqnarray}
p^{r}=\omega-q_e\phi,
\label{eq4.14}
\end{eqnarray}

\noindent where $\phi$ is the electric potential at the event horizon given by Eq. (\ref{eq2.9}). It recovers the relation gotten by the absorption of the scalar particle. For the spin down, the calculation is parrel and the same relation (\ref{eq4.14}) is gotten.

\section{Thermodynamics and particles' absorption}

Thermodynamics of black holes were deeply discussed through particles' tunnelling across event horizons. In this section, it is shown that the thermodynamics in the extended phase space of the five-dimensional black hole can be recovered by the absorptions of the scalar particle and fermion. The relation $p^{r}=\omega-q_e\phi$ is used in the derivation. Therefore, the particle can be either a scalar particle or a fermion. When a particle with energy and charge falls into the black hole, the energy and charge of the black hole increase. Let $f(M,Q,P,r_{+})=f(r_{+})$ represent the metric component of the original state of the black hole and $f(M+dM,Q+dQ,P+dP,r_{+}+dr_{+})$ be the metric component of the final state after the particle being absorbed. These metric components satisfy

\begin{eqnarray}
&& f(M+dM,Q+dQ,P+dP,r_++dr_+)-f(M,Q,P,r_+)=\nonumber\\
&&\left.\frac{\partial f(M,Q,P,r)}{\partial M}\right|_{r=r_+}dM + \left.\frac{\partial f(M,Q,P,r)}{\partial Q}\right|_{r=r_+}dQ+ \left.\frac{\partial  f(M,Q,P,r)}{\partial P}\right|_{r=r_+}dP \nonumber\\
&&+ \left.\frac{\partial f(M,Q,P,r)}{\partial r}\right|_{r=r_+}dr_+,
\label{eq5.1}
\end{eqnarray}

\noindent where

\begin{eqnarray}
\left.\frac{\partial f(M,Q,P,r)}{\partial M}\right|_{r=r_+}&=&-\frac{16\pi}{3\omega_{3}r_{+}^{2}},\quad \quad \left.\frac{\partial f(M,Q,P,r)}{\partial Q}\right|_{r=r_+}=\frac{32\pi^{2}Q}{3\omega_{3}^{2}r_{+}^{4}}, \nonumber\\ \left.\frac{\partial f(M,Q,P,r)}{\partial P}\right|_{r=r_+}&=&\frac{4\pi r_{+}^{2}}{3},\quad \quad\quad \left.\frac{\partial f(M,Q,P,r)}{\partial r}\right|_{r=r_+} =4\pi T.
\label{eq5.2}
\end{eqnarray}

\noindent Following Gwak's work, we assume that the final state is still a black hole. This implies $f(M+dM,Q+dQ,P+dP,r_{+}+dr_{+})= f(M,Q,P,r_{+})=0$. From Eqs. (\ref{eq5.1}) and (\ref{eq5.2}), the increase of the horizon radius takes the form

\begin{eqnarray}
dr_{+}=\frac{4p^{r}}{3\omega_{3}r_{+}^2T - 4\omega_{3}r_{+}^3P}.
\end{eqnarray}

\noindent Using the entropy relation $S=\frac{1}{4}\omega_3r_+^3$ yields

\begin{eqnarray}
dS=\frac{3p^{r}}{3T - 4r_+P},
\label{eq5.3}
\end{eqnarray}

\noindent which shows that the increase of the entropy is related to the Hawking temperature, pressure and $p^r$. For the non-extremal black hole or the fixed cosmological constant, we readily get $dS>0$, which implies that the entropy increase when the particle is absorbed by the black hole. In the extremal and near-extremal cases, $dS<0$. This implies that the entropy decreases when the particle falls into the black hole. It violates the second law of thermodynamics which shows the entropy never decreases in the clockwise direction.

We associate the particle's energy and charge with the increases of the internal energy and charge of the black hole,

\begin{eqnarray}
\omega =dU, \quad q_e = dQ.
\label{eq5.4}
\end{eqnarray}

\noindent Since the cosmological constant is regarded as the thermodynamic pressure, the black hole mass is seen as the enthalpy. The relationship between the mass and internal energy is

\begin{eqnarray}
U=M-PV.
\label{eq5.5}
\end{eqnarray}

\noindent Combining Eqs. (\ref{eq5.3}), (\ref{eq5.4}) and (\ref{eq5.5}) yields

\begin{eqnarray}
dM=TdS+\phi dQ+VdP,
\label{5.6}
\end{eqnarray}

\noindent which is the first law of thermodynamics recovered by the absorptions of the scalar particle and fermion. This derivation is inevitable, since the final state of the black hole after the particle being absorbed was assumed to be still a black hole.

\section{Overcharging problem in the extremal and near-extremal black holes}

In this section, we test the WCCC in the extremal and near-extremal five-dimensional black holes by the absorptions of the scalar particle and fermion. Here the final states of the black holes are unknown. An effective way to test the WCCC is to check whether a event horizon exists when a particle with energy and large enough charge or angular momentum falls into a black hole. The event horizon is determined by the metric component $f(r)$. If the minimal value of $f(r)$ is negative or zero, the horizon exists. Otherwise, the horizon does not exist.

We use $f(M,Q,P,r_{0})$ to represent the metric component of the original state and $f(M+dM,Q+dQ,P+dP,r_{0}+dr_{0})$ to denote the metric component of the final state, where $r_0$ is the location of the minimal value. The relation between these two metric components is

\begin{eqnarray}
&& f(M+dM,Q+dQ,P+dP,r_0+dr_0)= f(M,Q,P,r_0)+ \left.\frac{\partial f(M,Q,P,r)}{\partial r}\right|_{r=r_0}dr_0\nonumber\\
&&\left.\frac{\partial f(M,Q,P,r)}{\partial M}\right|_{r=r_0}dM + \left.\frac{\partial f(M,Q,P,r)}{\partial Q}\right|_{r=r_0}dQ+ \left.\frac{\partial  f(M,Q,P,r)}{\partial P}\right|_{r=r_0}dP ,
\label{eq6.1}
\end{eqnarray}

\noindent where

\begin{eqnarray}
f(M,Q,P,r_0)&=& k-\frac{M}{r_0^2} \frac{16\pi}{3\omega_3}+\frac{Q^2}{r_0^4} \frac{16\pi^2}{3\omega_3^2}+\frac{4}{3}\pi r_0^2P,\nonumber\\
\left.\frac{\partial f(M,Q,P,r)}{\partial r}\right|_{r=r_0} &=&0,  \quad\quad\quad\quad \left.\frac{\partial f(M,Q,P,r)}{\partial P}\right|_{r=r_0}=\frac{4\pi r_{0}^{2}}{3}, \nonumber\\
\left.\frac{\partial f(M,Q,P,r)}{\partial M}\right|_{r=r_0}&=&-\frac{16\pi}{3\omega_{3}r_{0}^{2}},\quad  \left.\frac{\partial f(M,Q,P,r)}{\partial Q}\right|_{r=r_0}=\frac{32\pi^{2}Q}{3\omega_{3}^{2}r_{0}^{4}},
\label{eq6.2}
\end{eqnarray}

\noindent For the extremal black hole, its Hawking temperature is zero. The location of the event horizon coincides with that of the minimal value, namely, $ r_{+}=r_{0} $. Thus,

\begin{eqnarray}
&&f(M+dM,Q+dQ,P+dP,r_{0}+dr_{0})\nonumber\\
&&=f(M,Q,P,r_{+})+\frac{16\pi}{3\omega_{3}r_{+}^{2}}\left[dM -VdP -\phi dQ\right]\nonumber\\
&&= 0,
\label{eq6.3}
\end{eqnarray}

\noindent where $f(M,Q,P,r_{+})=0$ and the first law of thermodynamics for the extremal black hole were used to derive the second equal sign. It shows that the final state of the extremal black hole is still a extremal one with new mass and charge. Therefore, the extremal black hole can not be overcharged.

In the near-extremal case, $f(M,P,Q,r_{0}) <0 $. We order $r_{0}=r_{+}-\epsilon$, where $0<\epsilon\ll 1$. Eq. (\ref{eq6.1}) is reduced to

\begin{eqnarray}
&&f(M+dM,Q+dQ,P+dP,r_{0}+dr_{0})=f(M,Q,P,r_{0})+\mathcal {O}(\epsilon)\nonumber\\
&&-\frac{16\pi}{3\omega_{3}r_{+}^{2}}\left(1+\frac{4\epsilon}{r_+}\right)\frac{3Tp^r}{3T-4r_+P}+\frac{2\epsilon}{r_+}\left(\frac{16\pi}{3\omega_{3}r_{+}^{2}}dM-4\pi r_+^2dP\right) .
\label{eq6.4}
\end{eqnarray}

\noindent The near-extremity leads to $T\to 0$. Considering $r_+ \sim l \gg 1 \gg \epsilon$, we neglect the second term of the second line in the above equation. Thus we get $f(M+dM,Q+dQ,P+dP,r_{0}+dr_{0})<0$. It shows that the final state of the near-extremal black hole is still a near-extremal one with the new mass and charge.

Therefore, the extremal and near-extremal five-dimensional black holes can not be overcharged by absorbing the scalar particle or fermion with mass and large enough charge.

\section{Discussion and Conclusion}

In this paper, we investigated the overcharging problem and thermodynamics in the extended phase spaces of the five-dimensional spherically symmetric topological black holes by the absorptions of the scalar particle and fermion. In the investigation, the cosmological constant was regarded as a variable related to pressure and its conjugate quantity is the thermodynamic volume. The calculations of the absorptions of the scalar particle and fermion reduced to the same relation $p^{r}=\omega-q_e\phi$. This relation is full in consistence with that gotten by the Hamilton-Jacobi equation \cite{BG}. Maybe the reason is that the Hamilton-Jacobi equation can be derived by the Klein-Gordon equation and Dirac equation. This derivation is referred to Ref. \cite{EL}. We used this relation to discuss the thermodynamics and recovered the first law of thermodynamics. This result is inevitable because the final state of the black hole after the particles being absorbed was assumed to be still a black hole. But the second law of thermodynamics was violated in the near-extremal and extremal black holes. The event horizon is determined by the metric component $f(r)$. To check the existence of the horizons at the final stages of the extremal and near-extremal black holes, we estimate the minimal values of the metric component. For the extremal black hole, its minimal value is zero, which shows that the final state is still a extremal black hole. For the near-extremal black hole, its minimal value is a negative value, which corresponds to a near-extremal black hole with new mass and charge. Therefore, the extremal and near-extremal black holes can not be overcharged.

\section{Acknowledgments}
This work is supported by the NSFC (Grant No. 11205125).

\end{document}